\documentclass[10pt,prb,aps,superscriptaddress,twocolumn,longbibliography]{revtex4-1}
\usepackage{amssymb,amsmath}
\usepackage{graphicx}
\usepackage{color}
\usepackage{natbib}
\usepackage{hyperref}
\usepackage[utf8]{inputenc}

\begin{document}

\newcommand{\e}{{\rm e}}
\newcommand{\rmi}{{\rm i}}
\renewcommand{\Im}{\mathop\mathrm{Im}\nolimits}
\newcommand{\iac}[1]{{\color{red}#1}}
\newcommand{\red}[1]{{\color{red}#1}}
\newcommand{\blue}[1]{{\color{blue}#1}}
\newcommand{\yellow}[1]{{\color{yellow}#1}}

\renewcommand{\cite}[1]{[\onlinecite{#1}]}
\newcommand{\cra}[1]{\hat{a}^{\dag}_{#1}}  
\newcommand{\ana}[1]{\hat{a}_{#1}^{\vphantom{\dag}}}         
\newcommand{\num}[1]{\hat{n}_{#1}}         
\newcommand{\ket}[1]{\left|#1\right>}      
\newcommand{\bra}[1]{\left<#1\right|}
\newcommand{\eps}{\varepsilon}      
\newcommand{\om}{\omega}      
\newcommand{\kap}{\varkappa}      

\newcommand{\spm}[1]{#1^{(\pm)}}
\newcommand{\skvk}[2]{\left<#1\left|\frac{\partial #2}{\partial k}\right.\right>} 
\newcommand{\skvv}[2]{\left<#1\left|#2\right.\right>}
\newcommand{\df}[1]{\frac{\partial #1}{\partial k}}
\newcommand{\ds}[1]{\partial #1/\partial k}

\title{Simulation of two-boson bound states \\ using arrays of driven-dissipative
coupled linear optical resonators}

\author{Maxim~A.~Gorlach}
\affiliation{ITMO University, Saint Petersburg 197101, Russia}
\email{m.gorlach@metalab.ifmo.ru}

\author{Marco Di Liberto}
\affiliation{INO-CNR BEC Center and Dipartimento di Fisica, Universit{\`a} di Trento, 38123 Povo, Italy}
\affiliation{Center for Nonlinear Phenomena and Complex Systems, Universit{\' e} Libre de Bruxelles, B-1050 Brussels, Belgium}

\author{Alessio Recati}
\affiliation{INO-CNR BEC Center and Dipartimento di Fisica, Universit{\`a} di Trento, 38123 Povo, Italy}

\author{Iacopo~Carusotto}
\affiliation{INO-CNR BEC Center and Dipartimento di Fisica, Universit{\`a} di Trento, 38123 Povo, Italy}

\author{Alexander N. Poddubny}
\affiliation{ITMO University, Saint Petersburg 197101, Russia}
\affiliation{Ioffe Institute, Saint Petersburg 194021, Russia}

\author{Chiara Menotti}
\affiliation{INO-CNR BEC Center and Dipartimento di Fisica, Universit{\`a} di Trento, 38123 Povo, Italy}

\begin{abstract}
We present a strategy based on two-dimensional arrays of coupled linear optical resonators to investigate the two-body physics of interacting bosons in one-dimensional lattices. In particular, we want to address the bound pairs in topologically non-trivial Su-Schrieffer-Heeger arrays. Taking advantage of the driven-dissipative nature of the resonators, we propose spectroscopic protocols to detect and tomographically characterize bulk doublon bands and doublon edge states from the spatially-resolved transmission spectra, and to highlight Feshbach resonance effects in two-body collision processes. We discuss the experimental feasibility using state-of-the-art devices, with a specific eye on arrays of semiconductor micropillar cavities.
\end{abstract}

\maketitle

\section{Introduction}\label{sec:Intro}

The Hubbard model is one of the most prominent theoretical models in physics. Since the early days of quantum condensed matter, it has been extensively applied to the study of the electronic properties of solids~\cite{Hubbard1963,Altland,Fradkin2013}. More recently, it has been extended to the description of ultracold atomic gases in optical lattices~\cite{Bloch:review2008}. In spite of its formal simplicity, the Hubbard model is able to capture most many-body properties of fermions and bosons in lattices, including superconductivity and magnetism in Fermi systems and, in the bosonic case, the superfluid to Mott insulator quantum phase transition~\cite{Fisher1989,Greiner2002}. One of its key features is that it can be straightforwardly extended to include additional bits of physics, e.g. different geometries and dimensionalities \cite{Lewenstein2012}, mixtures of several species~\cite{Soltan2011}, long-range interactions~\cite{Pfau2009,Ferlaino2016}. When non-trivial hopping coefficients that break time-reversal symmetry are introduced, the resulting effective magnetic field experienced by the atoms can lead to topological lattice models showing exotic many-body states~\cite{Sorensen2005,Moller:PRL2015,Vasic2015}.

In the last decade, the same Bose-Hubbard model has been exported to optics, so to describe arrays of coupled optical resonators showing sizeable $\chi^{(3)}$ optical nonlinearities that mediate effective interactions between photons~\cite{ICCC_RMP,angelakis2017quantum}. For strong nonlinearities, a number of theoretical works have investigated strongly correlated states of light, such as Mott insulators~\cite{Hartmann:Laser2008,noh2016quantum}. Special emphasis has been put on the intrinsic driven-dissipative nature of fluids of light~\cite{Biella:PRA2017,Lebreuilly:PRA2017}, which crucially distinguishes such setups from cold atoms experiments.
First experimental studies of photonic Mott insulator states are now starting to appear~\cite{simon}. 

The framework becomes even more intriguing in topological photonics models realized by means of synthetic magnetic fields for light~\cite{Lu2014,Lu2016,Khanikaev17,Ozawa_RMP}. In combination with optical nonlinearities, this leads to the emerging field of nonlinear topological photonics. Already for moderate nonlinearities, peculiar soliton and vortex states have been anticipated~\cite{Lumer,Leykam2016,Hadad,Solnyshkov,Gulevich,Smirnova2018}, while the interplay of synthetic magnetism with strong interactions has been predicted to give fractional quantum Hall states of light~\cite{Cho:PRL2008,Umucalilar:PRL2012}. First experimental steps in this direction have been recently reported~\cite{Roushan:2016NatPhys}.

Without entering into the complexity of many-body problems, already at the two-body level the Bose-Hubbard model reveals one of the most profound effects of periodicity~\cite{Mattis1986}, namely the formation of {\it doublons}, i.e. bound pairs, even for repulsive interactions. After their first observation with cold atoms~\cite{Winkler}, the study of these peculiar two-particle states has developed into an active area of research from both the theoretical~\cite{Petrosyan2007,Valiente,Valiente2009,Pinto,Flach,Menotti} and the experimental point of view~\cite{Strohmaier2010,Preiss2015,Mukherjee,Greiner2017}. In addition to the investigation of few-body edge localized states in standard Bose-Hubbard models~\cite{Flach,Longhi}, a special attention has been devoted in the latest years to the two-body physics in lattice models displaying topologically non-trivial bands at the single-particle level, with the ambitious goal of shining light on the interplay of interactions and topology at the few-body level. 
In this sense, the topological configurations that have attracted most interest include the Su-Schrieffer-Heeger model (SSH)~\cite{Bello2016,DiLiberto,DiLiberto-EPJ,Gorlach-2017,Marques2017,Marques2018} and the two-leg ladder with flux \cite{Greiner2017} in 1D, and the Haldane~\cite{Salerno} and the Hofstadter~\cite{Lee1} models in 2D.
In these systems, the existence of two-body edge states that may or may not be topologically protected has been shown, as well as the formation of lattice molecules originating from the resonance of bound pairs with scattering states~\cite{Lin,DiLiberto,Gorlach-2017,Salerno}.

   \begin{center}
    \begin{figure}[b]
     \includegraphics[width=0.8\linewidth]{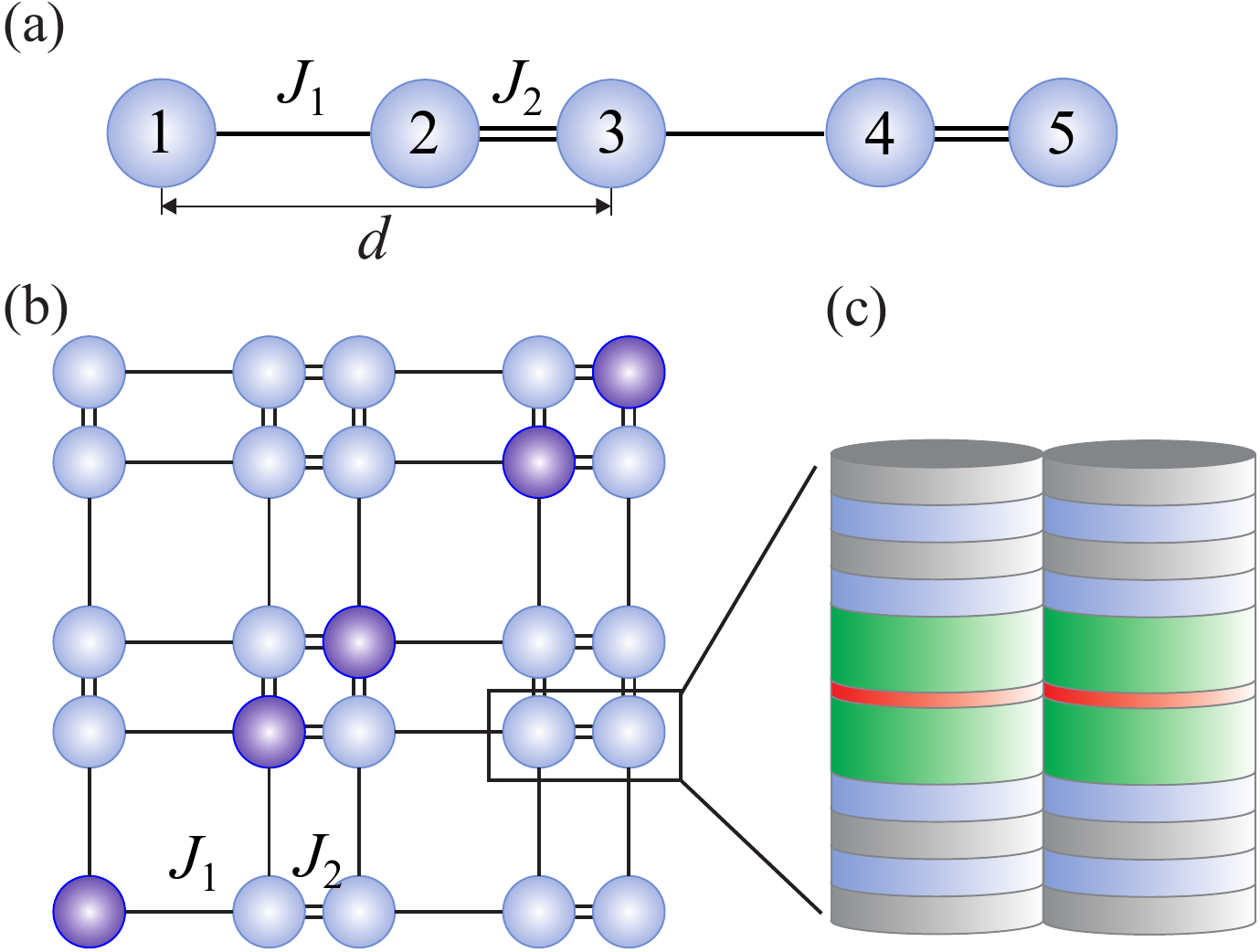}
    \caption{(a) Sketch of a dimerized lattice with superlattice unit cell $d$, providing the 1D Su-Schrieffer-Heeger (SSH) model described by the Bose-Hubbard Hamiltonian (\ref{BHHamiltonian}). (b) Two-dimensional classical setup emulating the two-particle quantum physics in the one-dimensional SSH model shown in (a). (c) Sketch of the polariton micropillar cavities that may be used to realize the two-dimensional lattice shown in (b) implementing the mapping of the one-dimensional two-body system shown in (a). 
    } 
    \label{fig:Pillars}
    \end{figure}
    \end{center}

A special feature of one-dimensional two-body systems is that they can be mapped onto two-dimensional single-particle systems, two-body interactions being included via external potentials, see Fig.~\ref{fig:Pillars}(a,b). The non-interacting nature of the final 2D system is then of utmost interest for optical implementations, since only linear optical elements are required.
In our previous works~\cite{DiLiberto,Gorlach-2017}, we proposed to simulate the two-body physics of a 1D SSH model by means of two-dimensional lattices of evanescently coupled optical waveguides. In the present work, we focus on two-dimensional arrays of coupled resonators. While the two schemes share the advantage of being based on photonic structures that are relatively easy to build, control and scale, their dynamics is completely different and provides access to different aspects of the same physics~\cite{ICCC_RMP,Ozawa_RMP}. 

On one hand, waveguides give access to the conservative temporal dynamics starting from a given initial input state. As one can evince from pioneering experiments~\cite{Schreiber:Science2012,Mukherjee}, these systems are an excellent platform to probe dynamical properties such as the effective doublon hopping rate, the beating phenomena related to the overlap of the input state with few eigenstates, and the real-time formation of Feshbach pairs. Doublon edge states may be instead difficult to identify because their small hopping rate makes a long time evolution needed to assess localization, a problem that could, for instance, be circumvented by the recently introduced state-recycling technique \cite{Mukherjee2018}.

On the other hand, the intrinsically driven-dissipative nature of the resonators appears favourable in view of spectroscopic studies of the two-body problem using the general approach reviewed in~\cite{ICCC_RMP}. Provided losses are kept at a moderate level, the different eigenstates of the Hamiltonian induce well-defined features in the reflection and transmission spectra: as we shall show in this work, resonant peaks can be associated with doublon bands and with doublon edge states of the equivalent 1D quantum non-dissipative problem.
A full tomography of the two-body scattering, bound and edge states can then be performed by varying the pump position in space and recovering their spatial profile from the emerging light. Moreover, the steady state intensity pattern allows one to highlight Feshbach resonance effects in two-body scattering processes via doublon states.

For the sake of concreteness, we will concentrate our attention on the specific case of arrays of semiconductor micropillar cavities like the ones sketched in Fig.~\ref{fig:Pillars}(c), whose potential to realize various models of topological physics has been validated by a number of recent experiments~\cite{Amo-PRL-2014,Sala,StJean}. Extension of our predictions to other platforms, such as the ring resonator arrays in Ref.~\cite{Hafezi:2013NatPhot}, is however straightforward.

The rest of the paper is organized as follows. The general two-body theory of doublon and doublon edge states in the interacting SSH model is reviewed in Sect.~\ref{sec:Doublons} and the basic concepts behind the 1D to 2D mapping are recalled in Sect.~\ref{sec:mapping}. After a short review of the effective 2D driven-dissipative system, Sect.~\ref{sec:results} presents our novel results: the spectroscopic signatures of doublon and doublon edge states in Sect.~\ref{sec:spectroscopy}, the eigenstate tomography in Sects.~\ref{Sect:tomography} and \ref{Sect:reconstruction}, and the evidence of a Feshbach resonance in~\ref{sec:Feshbach}. Conclusions and perspectives for future work are finally given in Sect.~\ref{sec:Discussion}.

\section{Two-body physics in the SSH model and 1D to 2D mapping: a brief summary}\label{sec:summary}

This section is devoted to a brief summary of the two-body physics in the interacting SSH model in the standard conservative case and to a review of the mapping of the one-dimensional two-body problem onto a two-dimensional single-particle one. Expert readers may skip this section and directly jump to Sect.~\ref{sec:results} where our new results in the presence of driving and dissipation are presented.

\subsection{Doublons and doublon edge states}
\label{sec:Doublons}

   \begin{center}
    \begin{figure}[!htbp]
      \includegraphics[width=1\linewidth]{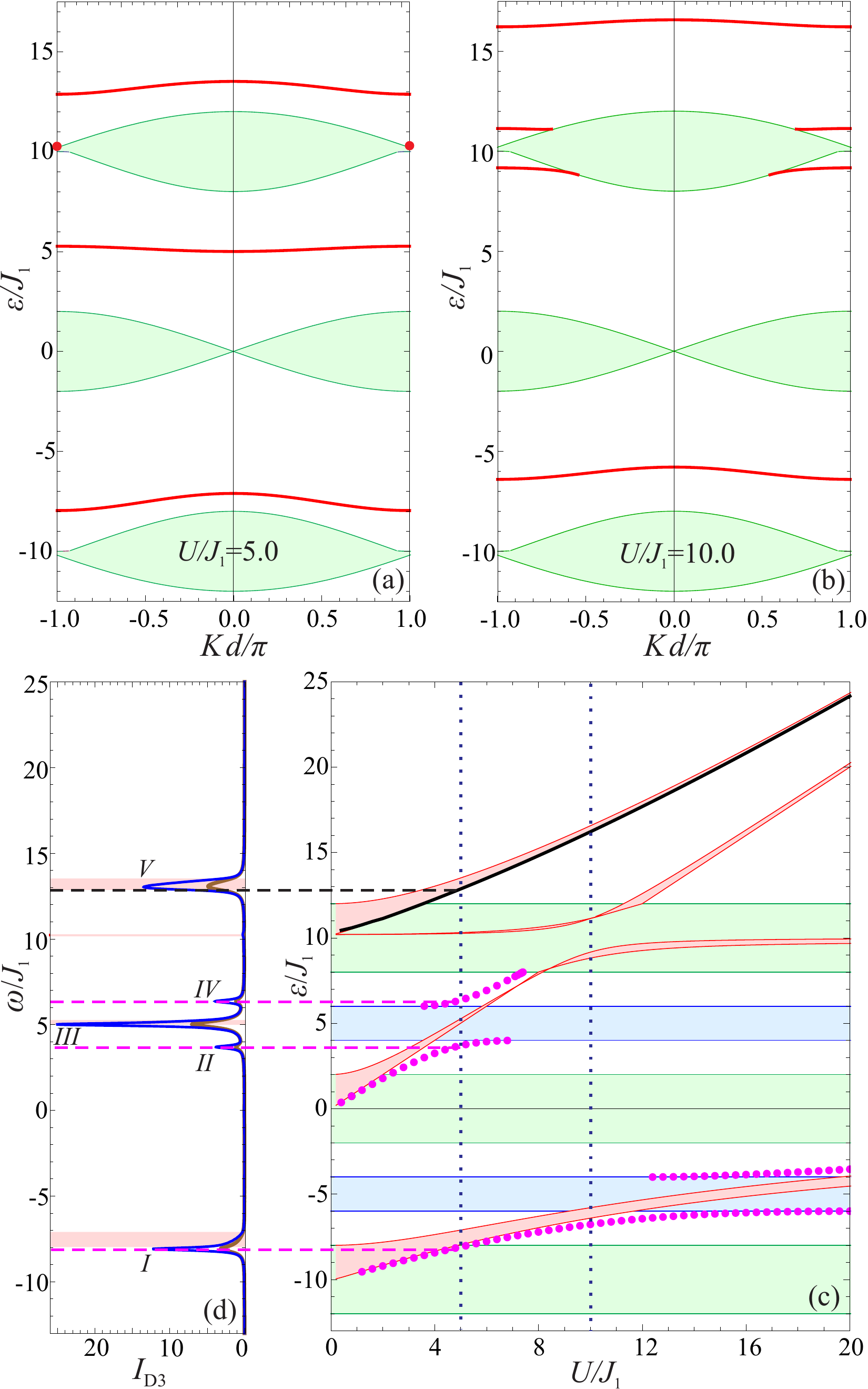}
      \caption{ 
        (a,b) Two-body excitation spectra in the SSH model for a system with periodic boundary conditions as a function of the center of mass momentum $K$, displaying scattering bands (green region) and bulk bound states (red bands), calculated for interaction strength (a) $U/J_{1}=5$ and (b) $U/J_{1}=10$. (c) Energy bands of two-boson excitations versus interaction strength $U$ for a system with open boundary conditions: scattering states (green), bulk bound states (red), single-particle edge states (blue), doublon edge states (magenta and black)~\footnote{To represent the bulk doublon bands, we considered the results for periodic boundary conditions. In such case, as shown in Fig.~\ref{fig:summary}(a,b), doublon states for a given value of $K$ are well-defined if no scattering states are available at the same energy. As soon as the doublon energy band enters the scattering continuum (or the single-particle edge state band), eigenstates acquire a hybridized nature and the distinction between doublon and scattering states is washed out. Physically, this mixing is responsible for a finite lifetime of the doublon states, possibly followed by revivals in a spatially finite system~\cite{CCT4}.}. (d) Integrated steady-state intensity from the pillars on the three main diagonals $m=n$, $m=n\pm 1$ obtained in a driven-dissipative system of $21\times 21$ sites, pump position in $(m,n)=(2,2)$, for two values of dissipation  $\gamma/J_{1}=0.1$ (blue) and  $\gamma/J_{1}=0.2$ (brown). Peaks in the spectrum are found in correspondence of bulk and edge doublon modes.
}
    \label{fig:summary}
    \end{figure}
   \end{center}
%


We consider the one-dimensional (1D) Hermitian system with nearest-neighbor tunneling and on-site interactions sketched in Fig.~\ref{fig:Pillars}(a) and described by the extended Bose-Hubbard Hamiltonian~\cite{Essler}
%
\begin{align}
\label{BHHamiltonian}
  \hat{H}^{\rm{BH}}=
  &-\sum\limits_{\langle m,n\rangle}\,J_{mn}\,\cra{m}\ana{n}
  +\frac{U}{2}\sum\limits_m\, \hat{n}_m\,\left(\hat{n}_m-1\right) \nonumber \\
  &+ \sum\limits_m\,\omega_m\,\hat{n}_m  \:.
\end{align} 
Here, $\cra{m}$ and $\ana{m}$ are the creation and annihilation operators for bosons at site $m$, respectively, and $\hat{n}_m=\cra{m}\,\ana{m}$ is the local density operator. A variety of lattice geometries can be described just by tuning the energy offsets $\omega_m$, the tunneling amplitude $J_{mn}$ between the $\langle m,n\rangle$ nearest neighboring sites ($J_{nm}=J_{mn}^{*}$), and the on-site boson-boson interaction constant $U$. In particular, the Su-Schrieffer-Heeger (SSH) model is characterized by alternating hopping coefficients $J_{1,2}$, constant $U$ and constant $\omega_m$. For convenience of notation, we set $\omega_m=0$ and use it as energy reference. In what follows, we will focus our attention on the two-particle sector of the interacting SSH Hamiltonian (\ref{BHHamiltonian}). 


A comprehensive analysis of the two-boson physics in the SSH chain was reported in our recent publications~\cite{DiLiberto,Gorlach-2017}. We give here a short summary of the main results to prepare the reader to the driven-dissipative case that will be discussed in the following sections.

Fig.~\ref{fig:summary}(a,b) shows the two-body energy spectrum as a function of the center of mass momentum $K$ in a chain with periodic boundary conditions for a fixed ratio of the tunneling amplitudes $J_2/J_1=5$ and two different values of the interaction strength $U$. In both cases, the two-body spectrum displays unbound scattering states (green bands) and tightly bound states, the so-called {\em doublons} (red bands). By varying $U$, the middle doublon band splits and crosses the upper scattering continuum.

Fig.~\ref{fig:summary}(c) shows instead the two-body spectrum for an open chain as a function of the interaction strength $U$. We consider  a chain with an odd number of sites $L$, that terminates at position $\ell=1$ with a weak link and at position $\ell=L$ with a strong link (see Fig.~\ref{fig:Pillars}(a) with $J_2>J_1$). In this case, edge states are also visible, in particular {\em single-boson edge states} 
with one boson localized at the weak link edge and the other boson delocalized over the chain (blue bands), and {\em doublon edge states} with both bosons simultaneously localized at the same edge of the array, either the weak-link edge (magenta dots) or the strong-link edge (black thick line). 
Note that some of the edge states shown in the spectrum  lie very close to other bands and thus have long localization lengths on the order of few or tens of lattice sites. Furthermore, in the neighborhood of $U\sim 2 J_2$  where the middle doublon band (red band) enters the upper scattering continuum (green band), hybridization between scattering and bound states occurs. A similar hybridization takes place for $U\sim J_2$ between the middle bulk doublon band and the upper single-particle edge states (blue band).

\subsection{Mapping of the two-body problem in 1D onto a single-particle problem in 2D}
\label{sec:mapping}

In view of the mapping of the 1D two-particle problem onto a 2D single-particle one, it is convenient to write the wavefunction in the first-quantized form
\begin{equation}\label{Psi}
\ket{\psi}=\sum\limits_{m,n}\,c_{mn}\,\ket{m,n}\;\;\;{\rm with}\, \sum_{m,n}|c_{mn}|^2=1\,,
\end{equation}
where $\ket{m,n}$ represents the state of two distinguishable particles, with the first particle at position $m$ and the second at position $n$. The indistinguishable and bosonic nature of the particles is taken into account by imposing the symmetry condition $c_{mn}=c_{nm}$.

The eigenvalue problem for Hamiltonian~\eqref{BHHamiltonian} can then be recast in terms of the set of linear equations for the coefficients $c_{mn}$
\begin{equation}
 \omega\,c_{mn} =\sum\limits_{m',n'} H_{mn,m'n'}\,c_{m'n'}
\label{CoupledMode00}
\end{equation}
with 
\begin{eqnarray}
H_{mn,m'n'}&=& \left(\omega_m+\omega_n+U\delta_{mn}\right)\delta_{mm'}\delta_{nn'} + \label{Hmnmn} \label{CoupledMode0}
\\
\nonumber &-&  \sum_{\eta=\pm 1} (J_{m m'} \delta_{m',m+\eta}\delta_{nn'} + J_{n n'}\delta_{mm'} \delta_{n',n+\eta}).
\end{eqnarray}
Coefficients $c_{mn}$ can thus be interpreted as the wavefunction of a single particle in a two-dimensional lattice of coordinates $m,n$, where $H$ is the Hamiltonian ruling its dynamics~\cite{Longhi:OL2011,Krimer:PRA2011,Schreiber:Science2012,Mukherjee}.

For the specific SSH model considered in this work, this mapping amounts in studying the dynamics of a single particle in the 2D lattice represented in Fig.~\ref{fig:Pillars}(b) and already introduced in our previous works~\cite{DiLiberto,Gorlach-2017}. All lattice sites have constant energy $\omega_m+\omega_n=0$ except for the sites on  the main diagonal of the lattice characterized by a local potential energy offset $U$ accounting for the on-site interaction in the corresponding 1D system. The alternating hopping parameters in 1D are mapped onto a specific structure of weak and strong couplings, as indicated in Fig.~\ref{fig:Pillars}(b) by single or double connecting lines.

\section{2D driven-dissipative system}
\label{sec:results}

The central point of this work is to propose a spectroscopic method to experimentally highlight the different two-body states summarized in Sect.~\ref{sec:Doublons} using a 2D array of coupled linear optical resonators. In contrast to the conservative temporal evolution typical of the experimental implementations in arrays of coupled optical waveguides~\cite{Schreiber:Science2012,Mukherjee}, our proposal relies on the driven-dissipative nature of the cavity dynamics, whose steady state stems from a dynamical interplay of coherent pumping and dissipation~\cite{ICCC_RMP}.

While our results are fully general and do not depend on the specific platform used to implement the 2D array, for the sake of concreteness, we will focus our attention on the specific case of semiconductor micropillar cavities sketched in Fig.~\ref{fig:Pillars}(c). Such systems~\cite{Amo:review} were demonstrated to allow an extreme flexibility in the design of the lattice geometry, ranging from 1D chains of many cavities~\cite{Sala,Baboux:PRL2016} to 2D honeycomb lattices~\cite{Amo-PRL-2014}. 

Semiconductor micropillar cavities are typically designed to operate in the near infrared around $1.5\,$eV right below the semiconductor gap, so that the cavity photon mode can be strongly coupled to an exciton transition. Since our proposal does not involve any optical nonlinearity, different choices for the cavity photon frequency can however be made, e.g. to increase the cavity quality factor to very large values. In addition to the overall control of the cavity resonance frequency via the thickness of the cavity layer, a fine tuning can be obtained by varying the pillar diameter (typically in the $d\approx 3.0~\mu$m range) and thus the transverse confinement energy of the photon. The strength of the coupling can instead be controlled via the distance between neighboring pillars. Typical values for the coupling energy are in the $0.25\,$meV range for distances in the $\rho=2.4~\mu$m range~\cite{Amo:review}. 

The driven-dissipative nature of the cavities can be included in the theoretical model by adding new terms describing coherent pumping and dissipative losses to the equations of motion in Eqs.~(\ref{CoupledMode00},\ref{CoupledMode0}).
The steady-state under a monochromatic pumping of frequency $\omega$ is described by 
\begin{equation}\label{CoupledMode}
\omega\,c_{mn}=\sum\limits_{m',n'} H_{mn,m'n'}\,c_{m'n'}-i\,\gamma\, c_{mn}-i\kap\,E_{mn}\:,
\end{equation}
where $c_{mn}$ are the steady-state amplitudes of the field at site  $(m,n)$, $H_{mnm'n'}$ is the effective Hamiltonian of the closed system defined in Eq.~(\ref{Hmnmn}), $\gamma$ is the dissipation rate, $E_{mn}$ is the spatial amplitude profile of the incident monochromatic field driving the cavity modes with coupling constant $\kap$. In order to isolate the physics of interacting bosonic particles, care must be paid to ensure that the pumping is symmetric, $E_{mn}=E_{nm}$.

{As it was reviewed at length in~\cite{ICCC_RMP}, the steady-state of the driven-dissipative system strongly depends on the frequency and the spatial profile of the coherent drive. In particular, the field intensity shows marked peaks when the external field is on resonance with some isolated eigenmode of the undriven system and the spatial profile of the pump overlaps with the eigenmode structure. The peak linewidth is controlled by the dissipation rate $\gamma$: the narrower this is, the more spectrally separated are the peaks from the surrounding features. In the next section, we will show how a suitable choice of the pump profile allows one to selectively address the different classes of states of the interacting two-body problem, including doublon and doublon edge states. Detailed information on their spatial structure can then be extracted, e.g., from a real-space image of the reflected or transmitted light.}

\subsection{Spectroscopy of doublons and doublon edge states}
  \label{sec:spectroscopy}

   \begin{center}
    \begin{figure*}[!ht]
            \includegraphics[width=1\linewidth]{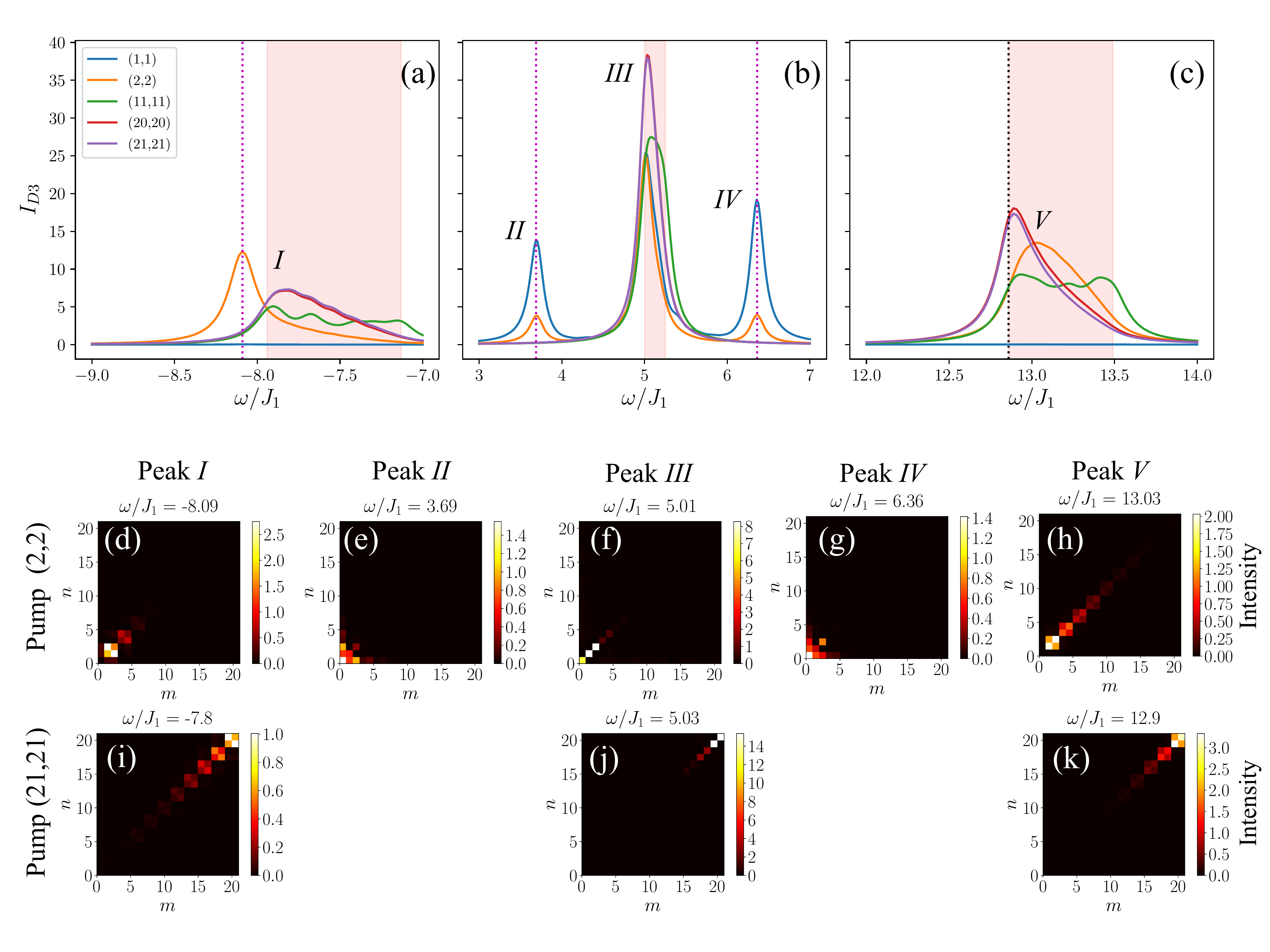}
\caption{(a,b,c) Spectroscopic analysis of the near-to-diagonal steady-state intensity $I_{D3}$ as a function of the pumping frequency $\omega$ for $U/J_1=5$, $\gamma/J_1=0.1$ and $L=21$. The relevant energy ranges are selected in panels (a) to (c) to highlight peak $I$, peak $II$, $III$, $IV$ and peak $V$, respectively. The different line colors indicate different pumping position, as detailed in the legend inset.
  The vertical magenta (black) dashed lines indicate the energy of the doublon edge states at the weak (strong) -link edge in the non-dissipative 1D chain.
  The red shaded areas indicate the energy range of the bulk doublon bands in the non-dissipative system.
  (d-k) Steady-state intensity profile for pumping in site $(2,2)$ (d-h) and for pumping in site $(L,L)$ (i-k). The pumping frequency is indicated in the different panels.}
\label{fig:spectroscopy}
    \end{figure*}
    \end{center}

We start by focusing our analysis on doublons, namely bound states where the two particles are strongly localized in the relative coordinate, which correspond to a large intensity close to the main diagonal of the 2D array $m=n$ and its neighbouring diagonals $m=n\pm1$ . Of particular interest are the doublon edge states, where -- with $L$ odd -- both particles are also simultaneously localized at one of the edges of the chain, corresponding to one corner of the 2D array, either near $m=n=1$ (\emph{weak-link corner}) or $m=n=L$ (\emph{strong-link corner}). In order to highlight the presence of these states, we consider a coherent pump localized on a single site of the main diagonal $m=n$ of the 2D array and look at the steady-state total intensity $I_{D3}$ summed over all pillars on the main diagonal $m=n$ and next to it $m=n\pm1$. Figure~\ref{fig:summary}(d) shows the spectrum of $I_{D3}$ as a function of the driving  frequency for a pumping position $(m,n)$ in the vicinity of the weak-link corner and two different values of the dissipation coefficient $\gamma=0.1J_1$ and $0.2J_1$. In both cases, the spectrum displays a series of marked and well isolated peaks. As expected, the spectral features are more evident for smaller dissipation $\gamma=0.1J_1$ (blue line), which is however close to the state-of-the-art values of present-day samples~\cite{Amo:review}.
   
The physical origin of the peaks can be inferred by comparing their position with the energy of the two-body eigenstates displayed in Fig.~\ref{fig:summary}(c): dashed horizontal lines indicate the energy of the doublon edge states, while the red shaded areas indicate the energy of the bulk doublon bands. However, a detailed analysis of the behaviour of the peaks as a function of pumping position is useful to discern between different types of bound states, namely between bulk and edge bound states.
To this aim, we zoom into the three specific frequency ranges corresponding to peaks $I$ to $V$ in Fig.~\ref{fig:summary}(d). For each peak, in Fig.~\ref{fig:spectroscopy}(a-c), we plot the evolution of the spectrum when the pump position is moved from one corner to the opposite one.

When pumping  close to the weak-link corner, specifically in $(2,2)$, peak $I$ emerges as a very marked feature at energy $\omega/J_1 \approx -8.1$, just below the bound state band [Fig.~\ref{fig:spectroscopy}(a)]. By moving the pumping position towards the center of the lattice, peak $I$ becomes smaller in amplitude and a broader structure develops in correspondence with the bulk bound states. Moving the pump towards the opposite corner $(L,L)$, a single broad peak is left. The fact that no excitation at all is obtained when the first site $(1,1)$ is pumped is due to the spatial structure of the underlying eigenstates. 

The steady-state intensity distributions $|c_{mn}|^2$ for the two pumping schemes, pumping position $(2,2)$ at pump frequency $\omega/J_1=-8.1$ and pumping position ($L,L$) at frequency $\omega/J_1=-7.8$, corresponding to the maxima of the obtained spectra, are shown respectively in Fig.~\ref{fig:spectroscopy}(d,i): while the overall intensity distributions are relatively similar, the one obtained by pumping in $(2,2)$ and shown in panel (d) is much more localized, highlighting the edge-localized nature of the corresponding state that one is able to excite. 
Through this analysis, one can relate the peak at $\omega/J_1 = -8.1$ obtained when pumping in $(2,2)$ to a quite extended doublon edge state that provides a dominant contribution to the spectrum when the pump is located next to the weak-link corner. On the other hand, the peak at $\omega/J_1 = -7.8$ which dominates the spectrum when the pump is located away from this corner [panel (i)] is due to the doublon band. 

Peak $III$ presents a series of broad and structured features. A quantitative information about the underlying eigenstates can be extracted by comparing the steady state intensity profiles at the maxima for pumping at different positions. When pumping at a generic diagonal $(m,m)$ position in the bulk, bulk doublons are created in doublon band states that ballistically move away from the pumping position within a characteristic distance fixed by the ratio of group velocity over $\gamma$.

Peaks $II$ and $IV$ in Fig.~\ref{fig:summary}(d) are well separated from peak $III$ and are related to very well localized doublon edge states. The very good localization properties of these states, connected to the appreciable energetic separation from the neighboring bands of single-particle edge states and bulk doublon states, is confirmed by the fact that the peaks quickly disappear when the pump position is moved away from the weak-link corner, as clearly shown in Fig.~\ref{fig:spectroscopy}(b). A very localized steady-state intensity pattern clearly emerges when pumping in $(2,2)$, as shown Fig.~\ref{fig:spectroscopy}(e,g).

Peak $V$ in Fig.~\ref{fig:summary}(d) shares some common features with peak $I$.
When pumping close to a corner --- in this case the strong-link corner at $(L,L)$ --- a relatively marked structure is present, which disappears and then shifts at slightly higher energies when moving the pumping position towards the opposite corner $(2,2)$.
This seems to hint towards the doublon edge state localized at the strong-link edge of the 1D chain.
%
As done above, some more information is obtained by looking at the steady-state intensity profiles, where one can actually observe signatures of a better localized doublon state for pumping in $(L,L)$ at energy $\omega/J_1 = 12.9$, shown in Fig.~\ref{fig:summary}(k), compared to a similar but not as much localized intensity distribution for pumping in $(2,2)$ at energy $\omega/J_1 = 13.03$, shown in Fig.~\ref{fig:summary}(h).

    \begin{center}
    \begin{figure}[t]
      \includegraphics[width=1\linewidth]{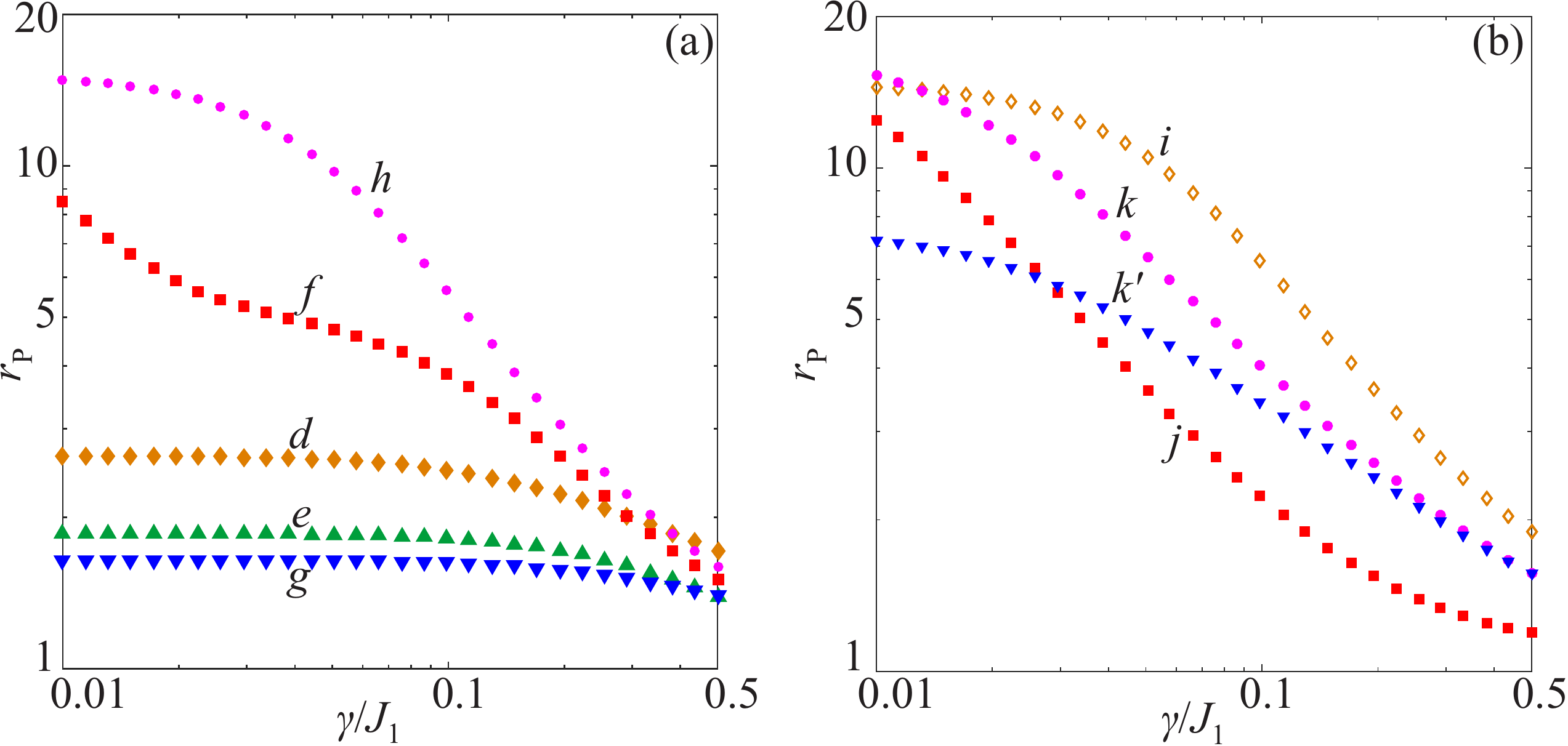}
      \caption{Localization properties of the steady-state intensity distributions as a function of $\gamma$; the localization properties are quantified by $r_p=\sqrt{\langle (x-m)^2+(y-n)^2 \rangle}$, depending on the pumping position $(m,n)$. (a) Weak-link localization obtained when pumping in $(m,n)=(2,2)$ at the same frequencies as in Fig.~\ref{fig:spectroscopy}: (d -- yellow diamonds), (e,g -- green and blue triangles), (f -- red squares), (h -- purple circles); (b) Strong-link localization obtained when pumping in $(m,n)=(L,L)$ at the same frequencies as in Fig.~\ref{fig:spectroscopy}: (i -- yellow diamonds), (j -- red squares), (k -- purple circles). For comparison, we plot also results for (k' -- blue triangles) obtained when pumping at energy $\omega/J_1=12.86$ corresponding to the edge bound state of the non-dissipative system.
}
    \label{fig:gamma}
    \end{figure}
    \end{center}

    A clear-cut distinction between bulk and edge bound states can be obtained by looking at the intensity distribution for different values of $\gamma$ smaller than the bulk doublon bandwidth: the intensity distribution of the edge bound state is expected not to change when $\gamma$ is reduced (and the spectral feature correspondingly narrows); conversely, the general driven-dissipative theory for the ballistic propagation in band states~\cite{ICCC_RMP}, predicts the propagation to occur up to a distance fixed by the ratio of group velocity over decay rate $\gamma$, corresponding to a spatial extension of the intensity distribution growing as $\gamma^{-1}$ when a doublon band is excited.

    These two typical behaviours emerge from the quantity $r_p=\sqrt{\langle (x-m)^2+(y-n)^2 \rangle}$ plotted in Fig.~\ref{fig:gamma}(a,b). We can identify cases (d,e,g) as the excitation of doublon edge states and cases (f,h,i,j) as the excitation of bulk doublon states. We have in fact checked that the saturation of curves (f,h,i,j) to the values around 15 in the very low $\gamma$ limit is only due to the finite size of the system $L=21$. On the other hand, the saturation for large $\gamma$ occurs when losses are large enough to hinder propagation, at values corresponding to one or two lattice sites.
    
    Case (k), which is obtained by pumping in $(L,L)$ site at frequency $\omega/J_1=12.9$, may seem to show the same behaviour predicted for bulk doublon states. However, by slightly lowering  the pumping frequency to $\omega/J_1=12.86$, so to exactly match the doublon edge state frequency predicted in the non-dissipative system, the behaviour changes in a dramatic way, as shown by case (k') in  Fig.~\ref{fig:gamma}(b). The saturation value for $r_p$ at small $\gamma$ is noticeably lower, highlighting the presence of an edge doublon state with very long localization length. 

In the next Subsect.~\ref{Sect:tomography} we will show how, for all the cases discussed above, important additional and more quantitative information can be obtained by performing a tomographic analysis of the eigenstates excited by a given pumping frequency.

    \begin{center}
    \begin{figure}[b]
      \includegraphics[width=1\linewidth]{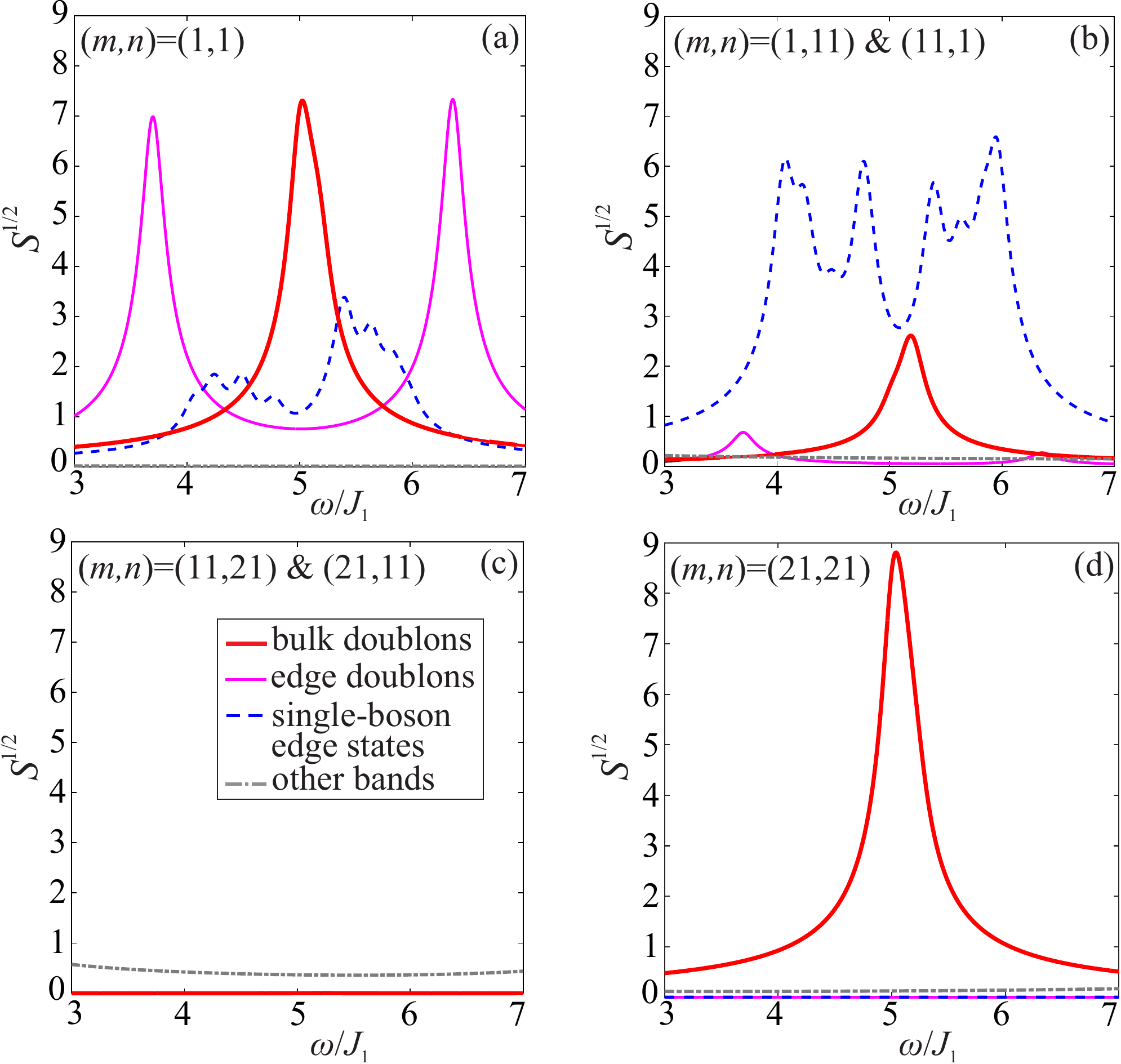}
      \caption{Overlap of the driven-dissipative system stationary state with bulk doublons (red solid lines), doublon edge states (solid magenta lines), single-boson edge states (dashed blue lines) for pumping frequency in the range $3<\omega/J_{1}<7$. The overlap with the rest of the states is shown by the dot-dashed gray lines and turns out to be negligible for those parameters. The calculations are performed for a system of $21\times 21$ sites, $U/J_{1}=5$,  $J_{2}/J_{1}=5$, $\gamma/J_{1}=0.1$ and pumping positions (a) $(m,n)=(1,1)$; (b) $(m,n)=(1,11)\&(11,1)$; (c) $(m,n)=(11,21)\&(21,11)$; (d) $(m,n)=(21,21)$.
       }
    \label{fig:Overlap}
    \end{figure}
    \end{center}


\subsection{State tomography}
\label{Sect:tomography}

To gain further insight about the obtained steady-state patterns and their relation with the eigenmodes of the non-dissipative system, we project the stationary state $\ket{\Psi}$ of the driven system onto the orthonormal basis of the eigenmodes $\ket{\phi_i}$ of the corresponding lossless system.
Since  $\ket{\Psi}$ is not necessarily normalized, the total overlap $\sum_i\,|\skvv{\phi_i}{\Psi}|^2$ is not fixed (in particular not equal to unity) and can be interpreted as the effective coupling of  a certain driving scheme to the underlying eigenstates of the system.
We define the overlap with a given subset of states $B$, e.g. scattering states, doublon or edge states, as
\begin{eqnarray}\label{OverlapDefinition}
S=\sum\limits_{i\in B}\,|\skvv{\phi_i}{\Psi}|^2\:\:,
\end{eqnarray}
where the sum is performed over all states $i$ belonging to the given subset.

As a simple but illuminating example, we analyze the composition of the steady state for $U/J_1=5$ and driving frequency in the range $\omega/J_1 \in [3,7]$. In this range of energies, one finds a pair of doublon edge states [magenta dots in Fig.~\ref{fig:summary}(c)], a single-particle edge-state band (blue region) and a narrower bulk doublon band (red region).
In Fig.~\ref{fig:Overlap}, we show the overlap with such eigenstates
as a function of driving frequency and for four different pumping positions. 
The contribution from the remaining states is shown by the gray dot-dashed line and turns out to be almost negligible for those parameters.

Fig.~\ref{fig:Overlap} clearly shows that, depending on the choice of the pumping point and driving frequency, the stationary state can have dominant overlap with different eigenstates of the non-dissipative system.
When pumping in corner $(1,1)$ as in Fig.~\ref{fig:Overlap}(a),
one mainly excites with a very precise frequency selectivity the doublon-edge states (magenta solid line) or the bulk doublon band (red solid line).
The overlap with such doublon states exhibits a truly resonant behavior with a linewidth related to the loss rate. For frequencies at which neither bulk nor edge doublon states are present, one observes a slighlty increased excitation of single-particle edge states (blue dashed line).

Moving the pumping profile along the weak-link border, e.g. at $(m,n)=(1,11)\&(11,1)$, the excitation of single-particle edge states (blue dashed line) is strongly enhanced as shown in Fig.~\ref{fig:Overlap}(b). Such states, in fact, are charaterized by an intensity distribution mostly localized along the two borders $(1,n)$ and $(m,1)$ of the $2D$ lattice corresponding to a single particle localized at the weak-link edge of the 1D chain.

In the center of the frequency range, one still observes a significant excitation of the bulk doublon band (red solid line). One can observe that the red curve is much less affected than the magenta line when moving the pumping position away from the corner along the weak-link borders of the 2D array.  Our interpretation is that for pumping positions on the weak-link borders of the 2D array, one actually accesses single-particle edge states that are strongly hybridized with bulk doublon states. To confirm such interpretation, we pump at position $(m,L)$ and $(L,n)$ located on the opposite strong-link borders, where such hybridisation phenomenon can not take place as no single particle edge states are present. As expected, none of the selected states is indeed excited [see Fig.~\ref{fig:Overlap}(c)]. Moving further to the opposite corner $(L,L)$ of the 2D array, no strong-link edge state (neither single nor two-particle) is excited, but a very strong overlap with bulk doublon states is recovered, as expected and shown in Fig.~\ref{fig:Overlap}(d).

\subsection{Eigenmode reconstruction}
\label{Sect:reconstruction}

    \begin{center}
    \begin{figure}[t]
      \includegraphics[width=1\linewidth]{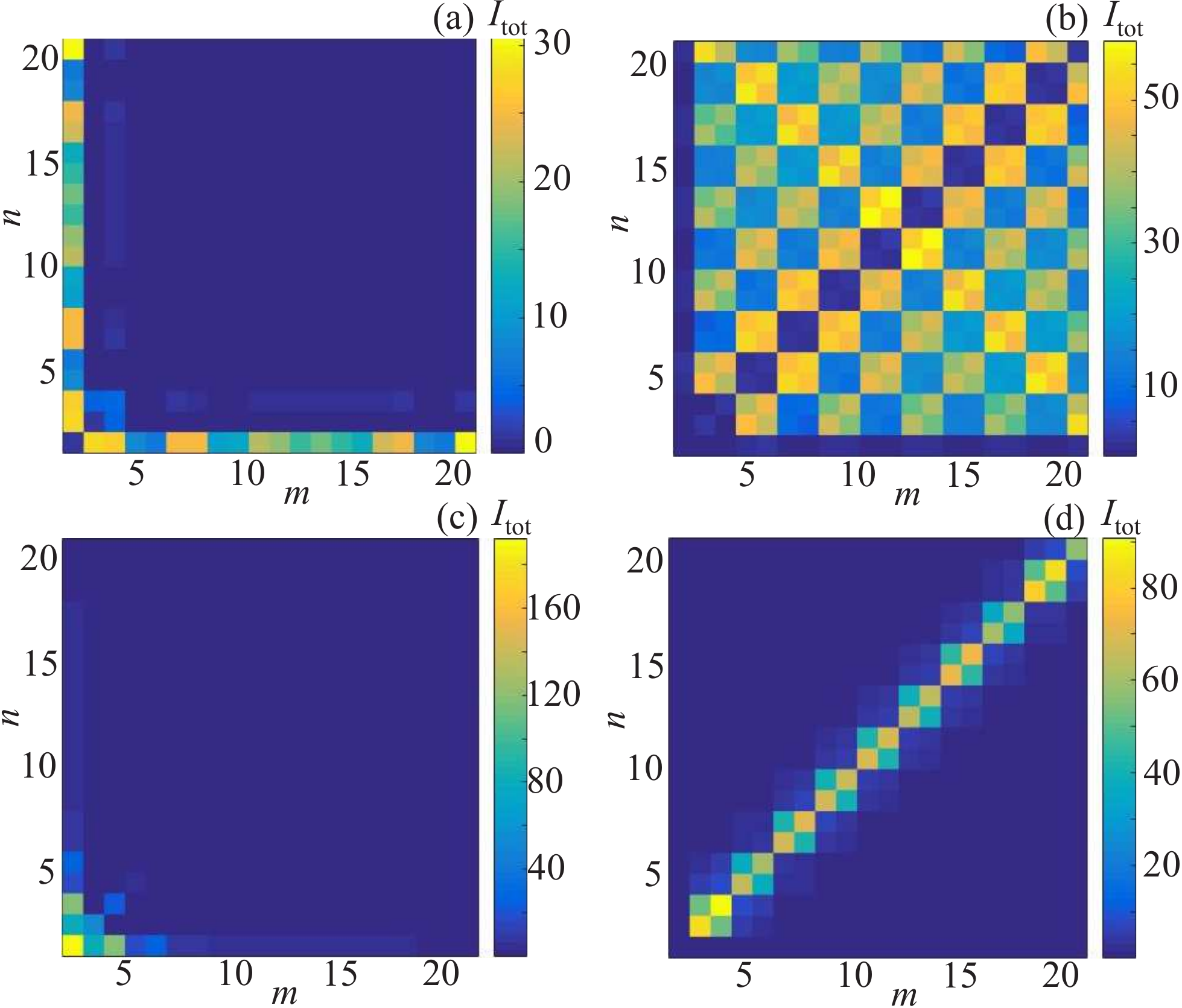}
      \caption{State tomography in an array of $21\times 21$ pillars for $J_{2}/J_{1}=5$, $U/J_{1}=5$, $\gamma/J_{1}=0.1$. The color maps represent the total integrated steady-state intensity as a function of pumping point $(m,n)$. Depending on the driving frequency, one can visualize different types of two-boson states:
        (a) single-boson edge state for $\omega/J_{1}=-5$;
        (b) scattering state of two bosons for $\omega/J_{1}=0$;
        (c) doublon edge state for $\omega/J_{1}=3.68$;
        (d) bulk doublon state for $\omega/J_{1}=13.25$.}
    \label{fig:OverlapMap}
    \end{figure}
    \end{center}

While the results of the last subsection confirm the capability of the pump to selectively excite different classes of eigenmodes depending on its position and frequency, they are not easily amenable to experimental implementation as they require a complete phase-sensitive information on the field amplitude and some {\em a priori} knowledge of the states. In this section, we propose an alternative scheme that is able to reconstruct
the intensity profile of the dominant eigenmode by mapping the total steady-state intensity as a function of pumping position.

The potential of this scheme is illustrated in the colorplots displayed in Fig.~\ref{fig:OverlapMap}: in each panel, the pump frequency and the lattice parameters are kept fixed, and the total intensity in the array is shown as a function of the pump location. In particular, panel (a) refers to a pump frequency exciting a state with a single particle bound to an edge and the second one being free to move. Panel (b) shows the excitation of scattering states with both particles free to propagate. Panel (c) shows the excitation of doublon edge states and, finally, panel (d) illustrates the excitation of a doublon band. 

A key advantage of this method is that it automatically performs a sum over all states within a bandwidth set by the loss rate $\gamma$. When one deals with a single isolated discrete state [e.g. the bound edge states of (c)] or an isolated narrow band of states [e.g. the doublon band of (d)], the sum is complete and the spatial distribution is a faithful representation of the states under examination. If one is dealing with a broader band, artifacts may instead appear due to the selective population of states within a limited bandwidth, resulting in the spatial oscillations that are well visible in (a,b).

\subsection{Feshbach resonance}
\label{sec:Feshbach}

    \begin{center}
    \begin{figure}[b]
    \includegraphics[width=0.95\linewidth]{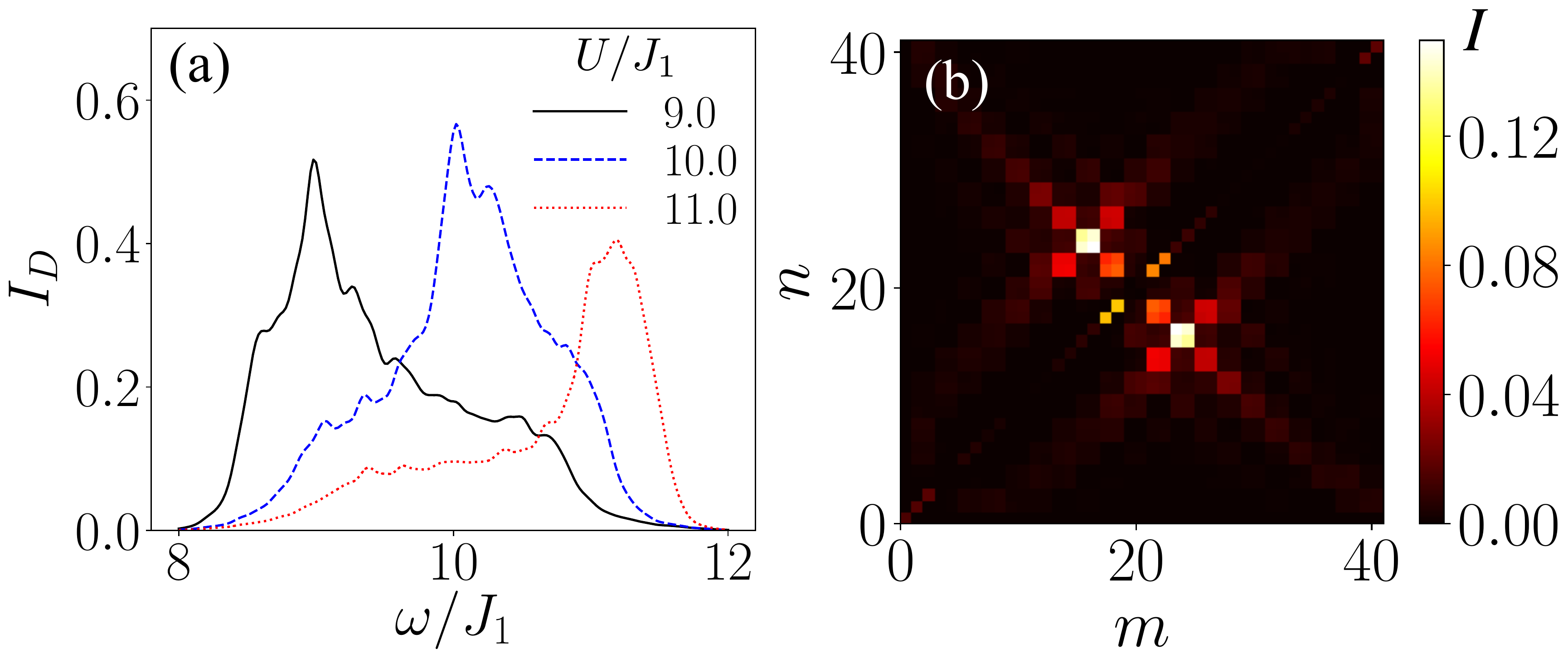}
    \caption{Demonstration of Feshbach resonance in a driven-dissipative system.
      (a) Total field intensity on the diagonal pillars $(m,m)$ as a function of
      pumping frequency for three different values of interaction
      strength $U/J_1=9,10,11$.
      (b) Steady-state intensity profile calculated for $J_2/J_1=5$, $U/J_1=10$, $\omega/J_1=U$, $\gamma/J_1=0.1$ and pumping position $(16,25)\&(25,16)$ shown by the white squares.
      Excitation of Feshbach molecules is observed through a sizeable population on the diagonal pillars.
      The calculations are performed for an array of
      $41 \times 41$ sites in order to minimize finite size effects.}
    \label{fig:Feshbach}
    \end{figure}
    \end{center}

As it was highlighted in Ref.~\cite{DiLiberto} and summarized in Sect.~\ref{sec:Doublons}, for suitable values of $U$ (namely $U\sim 2 J_2$)  the middle band of bulk doublon states crosses the upper scattering continuum. As a consequence, the eigenstates acquire a mixed scattering and bound character. In analogy to Feshbach resonance physics, this allows for the creation of an intermediate bound state during the scattering of two colliding particles. 

We now illustrate how this physics can be addressed in the 2D driven dissipative system considered in this work. To this purpose, one needs to pump sufficiently far away from the diagonal (while of course maintaining the bosonic symmetry $E_{mn}=E_{nm}$) at frequency close to the resonance position [crossing between red and green bands in Fig.~\ref{fig:summary}(a-c)]. In this way, the analog of an initial state with a pair of spatially separated incident particles is generated. The resonant nature of the scattering process then shows up as an enhanced steady-state intensity on the main diagonal when the frequency matches the appropriate resonance value for a given interaction strength $U$, namely when $\omega \sim U$.

In particular, we consider interaction strengths fixed to $U/J_1=9,10,11$ and scan the pumping frequency in the range $\omega/J_1 \in [8,12]$, spanning the upper scattering continuum.
In this frequency  window, hybridization between scattering and bound states corresponds to a sizeable steady-state intensity on the diagonal sites when the incident wave is resonant with the doublon bound state. This effect is responsible for the resonant behaviour observed in Fig.~\ref{fig:Feshbach}(a).  This interpretation is further corroborated by the shift of the resonant peak as a function of the interaction strength, which matches the shift of the doublon state with $U$ visible in Fig.\ref{fig:summary}(c). 

To conclude, in Fig.~\ref{fig:Feshbach}(b), we present an example of the steady-state intensity distribution for a pump frequency at the peak of the resonance: stripes emerging from the pumped sites correspond to independent propagation of the two particles. The X-shaped arrangement of the stripes reflects the diamond-like shape of the iso-energy curve in $k$-space for pumping in the middle of the band of scattering states. Particle collisions occur when these stripes hit and convert into a localized feature along the main diagonal, which corresponds to bound doublon states that are populated during the collision process.

\section{Discussion and outlook}\label{sec:Discussion}

In this work, we have shown how the quantum physics of two interacting particles in a one-dimensional chain can be simulated using a two-dimensional array of coupled linear resonators. Information on the eigenstates of the quantum problem can be retrieved by purely classical means from the transmission and reflection spectra of the optical system. By tuning the frequency of the incident light, the spatial shape of the different eigenstates can be inferred from the spatial profile of the transmitted light and fully reconstructed from the dependence of the total transmitted intensity on the pumping location. 

While this idea applies to generic models, our attention was focused on a Su-Schrieffer-Heeger chain with two alternating  tunneling constants and  on-site interactions, which shows a relatively simple but already very rich physics. For instance, our proposed configuration allows one to recover fundamentally quantum objects such as repulsively bound boson pairs as well as to visualize Feshbach resonance features in the two-body scattering problem. 

While arrays of semiconductor micropillar cavities turn out to be a most promising platform for the experimental implementation of our proposal, our results straightforwardly extend to generic cavity arrays in the visible, infrared or even microwave region of the spectrum, as well as to the waves of different nature such as sound. 

A direct next step of this work will be to assess the promise of our scheme to characterize the two-particle physics in more complex one-dimensional lattices and to extend our study to the fermionic case. While next-neighbor or even longer range interactions are straightforwardly included via a potential energy shift that extends over one or more sites around the main diagonal, the Fermi statistics can be enforced by implementing spatially anti-symmetric pump profiles. On the long run, a far more ambitious challenge will be to make use of the synthetic dimension concept~\cite{Ozawa:2016PRA} to simulate two-body physics in two-dimensional lattices using an effectively four-dimensional lattice.

\section{Acknowledgments}

We acknowledge fruitful discussions with A. Amo and G. Salerno. This work was supported by the Russian Science Foundation (Grant No. 18-72-00102). M.A.G. and A.N.P. acknowledge partial support by the Foundation for the Advancement of Theoretical Physics and Mathematics ``Basis".  M.D.L., A.R., I.C. and C.M acknowledge funding from Provincia Autonoma di Trento, partly through the SiQuro project (``On Silicon Chip Quantum Optics for Quantum Computing and Secure Communications") and from the EU-FET Proactive grant AQuS, Project No.640800. M.D.L. also acknowledges support from the ERC Starting Grant TopoCold.

\bibliography{TopologicalLib}

\end{document}